\documentclass[12pt]{article}

\usepackage{amsmath,amsfonts,amssymb,amsthm,textcomp,mathrsfs,mathrsfs,bbm}
\usepackage{braket,marvosym}
\usepackage{mathtools,stmaryrd}
\usepackage[utf8]{inputenc}
\usepackage{graphicx}
\usepackage{color}
\usepackage[colorlinks=true,linkcolor=black,citecolor=black,plainpages=false,pdfpagelabels]{hyperref}
\usepackage{palatino}
\usepackage[T1]{fontenc}
\usepackage{fullpage}

\newtheorem{theorem}{Theorem}
\newtheorem{lemma}[theorem]{Lemma}

\newtheorem{definition}{Definition}

\newcommand*{\cN}{\mathcal{N}}

\newcommand*{\brackett}[3]{\left\langle #1 \right| \left. #2 \right. \left| #3 \right\rangle}

\newcommand*{\eye}{{\mathbbm{1}}}
\newcommand*{\mzero}{{\mathbf{0}}}

\newcommand*{\cR}{\mathcal{R}}
\newcommand{\beq}{\begin{equation}}
\newcommand{\enq}{\end{equation}}
\newcommand{\ketbra}[1]{\ket{#1} \bra{#1}}

\newcommand{\tr}{\mathrm{Tr} \, }
\newcommand{\trns}{\mathrm{Tr}}

\newcommand*{\renyi}{R\'{e}nyi }

\usepackage[colorlinks=true,linkcolor=black,citecolor=black,plainpages=false,pdfpagelabels]{hyperref}
\hypersetup{pdftitle={Template}}
\usepackage{color}

\begin{document}

\title{More on a trace inequality in quantum information theory}
\author{Naresh Sharma \\
Tata Institute of Fundamental Research \\
Mumbai 400005, India \\
\Letter: \href{mailto:nsharma@tifr.res.in}{nsharma@tifr.res.in}}
\date{\today}
\maketitle

\begin{abstract}
It is known that for a completely positive and trace preserving (cptp) map ${\cal N}$,
$\text{Tr}$ $\exp$$\{ \log \sigma$ $+$ ${\cal N}^\dag[\log \cN(\rho)$ $-\log {\cal N}(\sigma)] \}$
$\leqslant$ $\text{Tr}$ $\rho$ when
$\rho$, $\sigma$, ${\cal N}(\rho)$, and ${\cal N}(\sigma)$ are strictly positive.
We state and prove a relevant version of this inequality
for the hitherto unaddressed case of these matrices being nonnegative.
Our treatment also provides an alternate proof for the strictly positive case.
\end{abstract}

\section{Introduction}

For a density matrix $\rho$ and a nonnegative $\sigma$, the quantum relative entropy is defined as
\beq
S(\rho \| \sigma) \equiv
\left\{
\begin{array}{ll}
\tr \rho ( \log \rho - \log \sigma) & \text{if } \text{supp}(\rho) \subseteq \text{supp}(\sigma) \\
\infty & \text{otherwise}
\end{array}
\right.
\enq
We confine ourselves to finite dimensional Hilbert spaces in this paper.
For a cptp map $\cN$, the monotonicity of the relative entropy states that
\beq
S(\rho \| \sigma) \geqslant S[ \cN(\rho) \| \cN(\sigma) ].
\enq

Equality conditions for the monotonicity of the quantum relative entropy
has been a well-studied topic \cite{petz-equality-rentr-1986,petz-mono-1988,
ruskai-equality-2002,petz-mono-2003,equality-ssa-2004,anna-equality-2009,sharma-qinf-2012,carlen-lieb-2014}.

Another topic has been the strengthening of the monotonicity by giving nonnegative lower bounds to
the remainder term:
\beq
\Delta(\rho,\sigma,\cN) \equiv S(\rho \| \sigma) - S[ \cN(\rho) \| \cN(\sigma) ].
\enq
Clearly, this would imply that there is equality in the monotonicity only if
the lower bound is zero although it may not always imply a sufficient condition.

The Fidelity between nonnegative matrices $X$ and $Y$ is defined
as $F(X,Y)$ $\equiv$ \linebreak $\| \sqrt{X} \sqrt{Y} \|_1$.
For $\alpha \in (0,2] \backslash \{1\}$, density matrix $\rho$ and nonnegative $\sigma$,
the \renyi $\alpha$-relative entropy \cite{lennert-2013,wilde-2013} is defined as
\beq
S_\alpha(\rho \| \sigma) \equiv \frac{1}{\alpha-1} \log
\tr \left( \sigma^{\frac{1-\alpha}{2 \alpha}}
\rho \sigma^{\frac{1-\alpha}{2 \alpha}} \right)^\alpha, ~~
\alpha \in (0,2] \backslash \{1\}.
\enq
To include $\alpha = 1$, we take limits and drop the subscript.

There has been quite a bit of work done in obtaining lower bounds of the form:
\beq
\label{yae0}
\Delta(\rho,\sigma,\cN) \geqslant S_{\frac{1}{2}} \left[\rho \| \cR(\rho) \right] =
-2 \log F \left[ \rho, \cR(\rho) \right],
\enq
where $\cR$ is, possibly a rotated version of, a cptp map
(see Refs. \cite{berta-2014,sutter-2015,junge-mono-2015} and references therein).

If the goal is to arrive at the best lower bound on $\Delta(\rho,\sigma,\cN)$,
then there seems to be no reason to expect
that, if it is in terms of $\rho$ and $\cR(\rho)$, then that $\cR$ ought to be cptp. On the other hand,
if one is looking for physically meaningful interpretations, then one would like to impose
that $\cR$ ought to be cptp.

Ruskai's equality conditions can be easily seen to be in terms of a positive
non-linear map \cite{ruskai-equality-2002} given by
\begin{align}
\label{yae5}
\widetilde{\cR}_{\sigma,\cN}(\tau) & \equiv \exp \left\{ \log \sigma + \cN^\dag \left[ \log \cN(\tau)
- \log \cN(\sigma) \right] \right\},
\end{align}
where $\sigma, \cN(\tau), \cN(\sigma)$ are all strictly positive.
This map arises most naturally without any sophisticated operations on the remainder term.
Clearly, $\widetilde{\cR}_{\sigma,\cN}(\sigma) = \sigma$.
In Ref. \cite{ruskai-equality-2002}, Ruskai leaves open the question of proving or disproving
\beq
\label{yae11}
\tr \widetilde{\cR}_{\sigma,\cN}(\tau) \leqslant \tr \tau
\enq
and mentions that proving \eqref{yae11}
paves the way to an alternate proof of Ruskai's equality condition that
$\Delta(\rho,\sigma,\cN) = 0$ if and only if $\rho = \widetilde{\cR}_{\sigma,\cN}(\rho)$.
Datta and Wilde proved \eqref{yae11} when all the matrices involved are strictly positive
\cite{ineq-first-proof-2015}. (See also Ref. \cite{zhang-mono-2014} for some related open
questions some of which have been addressed in Ref. \cite{ineq-first-proof-2015}.)

In this paper, we make the following two contributions.
\begin{itemize}
\item We provide and prove a relevant generalized
version of \eqref{yae11} when the matrices could be nonnegative. Its relevance comes from
its applicability to the equality conditions for, and the strengthening of, the monotonicity of
relative entropy.
\item The treatment also provides an alternate proof of \eqref{yae11} when all the matrices
involved are strictly positive.
\end{itemize}

We shall employ the Golden-Thompson inequality proved independently by
Golden \cite{golden-1965} and Thompson \cite{thompson-1965},
and its generalization due to Lenard \cite{lenard-1972} and Thompson \cite{thompson-1972} that
states that for Hermitian matrices $Y$ and $Z$ and any unitarily invariant norm
$\interleave \cdot \interleave$,
\beq
\label{gen-gt}
\interleave \exp(Y + Z) \interleave \leqslant \interleave \exp(Y/2) \cdot \exp(Z) \interleave,
\enq
where $U \cdot W \equiv U W U^\dag$,
and the Golden-Thompson inequality follows from the special case of $\| \cdot \|_1$.
We also employ the triple matrix generalization of the
Golden-Thompson inequality due to Lieb \cite{lieb-ssa-1973} (see also Ref. \cite{ruskai-equality-2002})
that states that for strictly positive $X,Y,Z$,
\beq
\label{t-gt}
\tr \exp(\log X - \log Y + \log Z) \leqslant \tr \int_0^\infty X (Y + t \eye)^{-1} Z (Y + t \eye)^{-1} dt.
\enq

\section{Main result}

This section gives the main result of this paper in the following theorem for the
hitherto unaddressed case of nonnegative matrices.
For the ease of reading,
a stripped-down version of this theorem for strictly positive matrices is provided in Appendix
\ref{alt-proof} and would constitute an
alternate proof of the Corollary 10 in Ref. \cite{ineq-first-proof-2015}.

Let $\Pi_X$ to be a projector on the
support of a nonnegative matrix $X$ and $\Pi_X^\bot \equiv \eye - \Pi_X$.
For a Hermitian matrix $Y$ with the spectral decomposition $Y = \sum_j \beta_j \ketbra{j}$,
we follow the $0 \log 0 = 0$ convention, and
for any projector $\Pi$ such that $\Pi \leqslant \Pi_Y$, 
we denote $\Pi \cdot \log(Y)$ to mean $\sum_{j: \beta_j > 0} (\log \beta_j) \Pi \ketbra{j} \Pi$.
The matrix with all entries as zero is denoted by $\mzero$.

We first provide an extended definition for $\widetilde{\cR}_{\sigma,\cN}$ and
\eqref{yae5} is obtained as a special case. The relevance of this definition is discussed in
Section \ref{discuss}.

\begin{definition}
For a cptp map $\cN$, nonnegative $\sigma$, $\tau$,
$\text{supp}(\tau)$ $\subseteq \text{supp}(\sigma)$, let
\beq
\Omega_{\sigma,\cN}(\tau) \equiv \Pi_\tau \cdot \left\{ \log \sigma + \cN^\dag \left[ \Pi_{\cN(\tau)} \cdot
\log \cN(\tau) - \Pi_{\cN(\sigma)} \cdot \log \cN(\sigma) \right] \right\}.
\enq
Since $\Pi_\tau = \tau^0$, $\Omega_{\sigma,\cN}(\tau)$ is a function of $\tau$.
Define
\begin{align}
\label{yae6}
\widetilde{\cR}_{\sigma,\cN} (\tau) & \equiv \lim_{\delta \downarrow 0}
\exp \left[ \Omega_{\sigma,\cN}(\tau) + (\log \delta) \Pi_\tau^\bot \right] = \Pi_\tau \cdot \exp \left[
\Omega_{\sigma,\cN}(\tau) \right].
\end{align}
\end{definition}

\begin{theorem}[Trace inequality for nonnegative matrices]
\label{theorem02}
For $\widetilde{\cR}_{\sigma,\cN}$ defined in \eqref{yae6},
$\tr \widetilde{\cR}_{\sigma,\cN}(\tau)$ $\leqslant$ $\tr \tau$.
\end{theorem}
\begin{proof}
For a $\delta > 0$, let $\widetilde{\cR}_{\sigma,\cN,\delta}(\tau) \equiv
\exp \left[ \Omega_{\sigma,\cN}(\tau) + (\log \delta) \Pi_\tau^\bot \right]$.
Let the partial isometry $V^{A \to BE}$ be an isometric extension of $\cN^{A \to B}$ and
$\hat{\Pi} \equiv \eye^{BE} - V V^\dag$. For any $\varepsilon > 0$, define
\begin{align}
\sigma_{1,\varepsilon}^{BE} & \equiv V \cdot \sigma + \varepsilon V \cdot \Pi_\sigma^\bot + \varepsilon \hat{\Pi},\\
\Upsilon_\varepsilon & \equiv \log \left[ \cN(\tau) + \varepsilon \Pi_{\cN(\tau)}^\bot \right] -
\log \sigma_{1,\varepsilon}^B, \\
\Theta_\varepsilon & \equiv \Pi_\tau \cdot \cN^\dag
\left[ -(\log \varepsilon) \Pi_{\cN(\tau)}^\bot + \log \sigma_{1,\varepsilon}^B -
\Pi_{\cN(\sigma)} \cdot \log \cN(\sigma) \right] \\
& = \Pi_\tau \cdot \cN^\dag \left\{ \log \sigma_{1,\varepsilon}^B -
\log \left[ \cN(\sigma) + \varepsilon^{1/2} \Pi_{\cN(\sigma)}^\bot \right] \right\},
\end{align}
where in the last equality, we first note using Lemma \ref{yal2} that $\Pi_{\cN(\sigma)}^\bot \leqslant
\Pi_{\cN(\tau)}^\bot$ and we invoke Lemma \ref{yal3} to claim
$\mzero = \Pi_\tau \cdot \cN^\dag(\Pi_{\cN(\tau)}^\bot) = \Pi_\tau \cdot \cN^\dag[\Pi_{\cN(\sigma)}^\bot]$.
We now have
\begin{align}
\label{yae7}
\tr \widetilde{\cR}_{\sigma,\cN,\delta^2}(\tau)
& = \tr \exp \left\{ \Pi_\tau \cdot \left[ \log \sigma + \cN^\dag(\Upsilon_\varepsilon) \right] + \Theta_\varepsilon
+ 2 (\log \delta) \Pi_\tau^\bot \right\} \\
& \leqslant \tr \exp \left\{ \Pi_\tau \cdot \left[ \log \sigma + \cN^\dag(\Upsilon_\varepsilon) \right] +
(\log \delta) \Pi_\tau^\bot \right\} \exp \left[ \Theta_\varepsilon + (\log \delta) \Pi_\tau^\bot \right] \\
\label{yae10}
& \leqslant \tr \exp \left\{ \Pi_\tau \cdot \left[ \log \sigma + \cN^\dag(\Upsilon_\varepsilon) \right] +
(\log \delta) \Pi_\tau^\bot \right\} \left\| \exp \left[ \Theta_\varepsilon + (\log \delta) \Pi_\tau^\bot \right] \right\|,
\end{align}
where the first inqquality follows using \eqref{gen-gt},
and the second inequality
follows from the fact that for nonnegative matrices $Y$ and $Z$, $\tr Y Z \leqslant \tr Y \| Z \|$,
where $\| \cdot \|$ is the operator norm. 
We attack the first term in \eqref{yae10}.
\begin{align}
\tr & \exp \left\{ \Pi_\tau \cdot \left[ \log \sigma + \cN^\dag(\Upsilon_\varepsilon) \right] +
(\log \delta) \, \Pi_\tau^\bot \right\} \\
& = \tr \exp \left\{ (\Pi_\tau V^\dag) \cdot \left( \log \sigma_{1,\varepsilon}^{BE} + \Upsilon_\varepsilon \otimes
\eye^E \right) + (\log \delta) \, \Pi_\tau^\bot \right\} \\
& \leqslant \tr \exp \left( \log \sigma_{1,\varepsilon}^{BE} + \left\{
\log \left[ \cN(\tau) + \varepsilon \Pi_{\cN(\tau)}^\bot \right]
- \log \sigma_{1,\varepsilon}^B \right\} \otimes \eye^E \right) + \delta \, \tr \Pi_\tau^\bot \\
& \leqslant \tr \int_0^\infty \sigma_{1,\varepsilon}^{BE} \left\{ (t \eye^B + \sigma_{1,\varepsilon}^B)^{-1}
\left[ \cN(\tau) + \varepsilon \Pi_{\cN(\tau)}^\bot \right]
(t \eye^B + \sigma_{1,\varepsilon}^B)^{-1} \otimes \eye^E \right\} dt
+ \delta \, \tr \Pi_\tau^\bot \\
& = \tr \int_0^\infty (t \eye^B + \sigma_{1,\varepsilon}^B)^{-1} \sigma_{1,\varepsilon}^{B}
(t \eye^B + \sigma_{1,\varepsilon}^B)^{-1} dt \, \left[\cN(\tau) + \varepsilon \Pi_{\cN(\tau)}^\bot \right]
+ \delta \, \tr \Pi_\tau^\bot \\
\label{yae8}
& = \tr \tau + \varepsilon \, \tr \Pi_{\cN(\tau)}^\bot + \delta \, \tr \Pi_\tau^\bot,
\end{align}
where the first equality follows since for any $\varepsilon \neq 0$, $\log \sigma_{1,\varepsilon}^{BE}
= V \cdot \log \sigma + (\log\varepsilon) V \cdot \Pi_\sigma^\bot + (\log\varepsilon) \hat{\Pi}$, and hence,
$(\Pi_\tau V^\dag) \cdot \log \sigma_{1,\varepsilon}^{BE} = \Pi_\tau \cdot \log \sigma$,
and $\cN^\dag(\gamma) = V^\dag \cdot (\gamma \otimes \eye^E)$, the first inequality uses
Lemma \ref{yal1}, and the second inequality uses \eqref{t-gt}
and we note that we can apply it since $\sigma_{1,\varepsilon}^{BE}$, $\sigma_{1,\varepsilon}^{B}$,
and $\cN(\tau) + \varepsilon \Pi_{\cN(\tau)}^\bot$ are all strictly positive.

We now attack the second term in \eqref{yae10}.
Let $X^B \equiv \cN(\Pi_\sigma^\bot) + \trns_E \hat{\Pi}$, and so
$\sigma_{1,\varepsilon}^B = \cN(\sigma) + \varepsilon X^B$. Let
$W$ be the square root of the pseudo-inverse of $\cN(\sigma)$ and hence,
$W \cN(\sigma) W^\dag = \Pi_{\cN(\sigma)}$. We have
\begin{align}
\big\| & \exp \left[ \Theta_\varepsilon + (\log \delta) \Pi_\tau^\bot \right] \big\| \nonumber \\
& = \big\| \exp \left( \Pi_\tau \cdot
\cN^\dag \left\{ \log \sigma_{1,\varepsilon}^B - \log \left[ \cN(\sigma) +
\varepsilon^{1/2} \Pi_{\cN(\sigma)}^\bot \right] \right\} + (\log \delta) \Pi_\tau^\bot \right) \big\| \\
& = \left\| \exp \left[ (\Pi_\tau V^\dag) \cdot \left( \left\{ \log \sigma_{1,\varepsilon}^B -
\log \left[ \cN(\sigma) + \varepsilon^{1/2} \Pi_{\cN(\sigma)}^\bot \right] \right\}
\otimes \eye^E \right) + (\log \delta) \Pi_\tau^\bot \right] \right\| \\
& \leqslant \max \left[ \left\| \exp \left( \log \left\{ \left[ \cN(\sigma) + \varepsilon X^B \right] \otimes \eye^E \right\} -
\log \left\{ \left[ \cN(\sigma) + \varepsilon^{1/2} \Pi_{\cN(\sigma)}^\bot \right]
\otimes \eye^E \right\} \right) \right\|, \delta \right] \\
& \leqslant \max \left( \left\| \left\{ \left[ W + \varepsilon^{-1/4}  \cdot \Pi_{\cN(\sigma)}^\bot \right] \cdot
\left[ \cN(\sigma) + \varepsilon X^B \right] \right\} \otimes \eye^E \right\|, \delta \right) \\
& = \max \left\{ \left\| \Pi_{\cN(\sigma)} + \varepsilon W \cdot X_B + \varepsilon^{3/4} [\Pi_{\cN(\sigma)}^\bot
X_B W + W X_B \Pi_{\cN(\sigma)}^\bot] + \varepsilon^{1/2} \Pi_{\cN(\sigma)}^\bot \cdot X_B \right\|, \delta
\right\} \\
\label{yae9}
& \leqslant \max \left[ 1 + f(\varepsilon), \delta \right],
\end{align}
where the first inequality uses Lemma \ref{yal1},
and the second inequality uses \eqref{t-gt}, and in the last inequality, we have used the triangle
inequality and
$f(\varepsilon) \equiv \varepsilon \| W \cdot X_B \| + \varepsilon^{3/4} \| \Pi_{\cN(\sigma)}^\bot
X_B W + W X_B \Pi_{\cN(\sigma)}^\bot \| + \varepsilon^{1/2} \| \Pi_{\cN(\sigma)}^\bot \cdot X_B \|$.

Using \eqref{yae8} and \eqref{yae9} in \eqref{yae10}, we get
\beq
\tr \widetilde{\cR}_{\sigma,\cN,\delta^2}(\tau)
\leqslant \left[ \tr \tau + \varepsilon \tr \Pi_{\cN(\tau)}^\bot + \delta \tr \Pi_\tau^\bot \right]
\times \max \left[ 1 + f(\varepsilon), \delta \right],
\enq
and hence,
\beq
\tr \widetilde{\cR}_{\sigma,\cN}(\tau)
\leqslant \left[ \tr \tau + \varepsilon \tr \Pi_{\cN(\tau)}^\bot \right]
\left[ 1 + f(\varepsilon)\right].
\enq
Since this is true for all $\varepsilon > 0$ and $f(\varepsilon)$ decreases to $0$ as
$\varepsilon$ goes to $0$, the claim follows.
\end{proof}

\section{Discussion}
\label{discuss}

For completeness, we provide an alternate proof to Ruskai's necessary and
sufficient equality condition for the monotonicity along with its so-called strengthening.

Consider a density matrix $\rho$, nonnegative $\sigma$, $\text{supp}(\rho)$ $\subseteq \text{supp}(\sigma)$,
and a cptp map $\cN$.
Note that when $\text{supp}(\rho)$ $\subsetneq \text{supp}(\sigma)$, there is nothing to prove since
the quantities involved are infinite.
From Theorem \ref{theorem02},
$\Theta \equiv \tr \widetilde{\cR}_{\sigma,\cN}(\rho) \leqslant 1$.
Using the notation $0 \log 0 = 0$, we have
\begin{equation}
\Delta(\rho,\sigma,\cN) = \tr \rho [ \log \rho - \Omega_{\sigma,\cN}(\rho) ]
= \tr \rho [ \log \rho - \log \widetilde{\cR}_{\sigma,\cN}(\rho) ]
= S \Big[ \rho \Big\| \frac{\widetilde{\cR}_{\sigma,\cN}(\rho)}{\Theta} \Big] - \log \Theta,
\end{equation}
which is the sum of two nonnegative terms where, from the Klein's inequality,
the first term is zero if and only if
$\rho = \widetilde{\cR}_{\sigma,\cN}(\rho)/\Theta$ and the second term is zero if and only if
$\Theta = 1$. Putting these two together implies
$\Delta(\rho,\sigma,\cN) = 0$ if and only if $\rho = \widetilde{\cR}_{\sigma,\cN}(\rho)$.
We also have
\beq
\Delta(\rho,\sigma,\cN) \geqslant S_\alpha [ \rho \| \widetilde{\cR}_{\sigma,\cN}(\rho)], ~~~ \alpha \in (0,1],
\enq
which follows by invoking the monotonicity of $S_\alpha$ in $\alpha$.
In particular, choosing $\alpha = 1/2$ gives a bound of the form in \eqref{yae0}.
This holds regardless of whether Theorem \ref{theorem02} holds or not, but
it becomes meaningful because Theorem \ref{theorem02} makes the lower bound nonnegative.
This is just an example and several off-the-shelf
lower bounds to relative entropy may be used to get meaningful lower bounds to
$\Delta(\rho,\sigma,\cN)$.

\section{Conclusions and Acknowledgements}

We state and prove a relevant version of a trace inequality
in quantum information theory
for the case of nonnegative matrices that has not been addressed thus far.
In doing so, we also provide an alternate proof of that
trace inequality for strictly positive matrices.

After the completion of this work, the author became aware of
Datta and Wilde's result in
Ref. \cite{ineq-first-proof-2015} and thanks M. M. Wilde for
pointing him to Ref. \cite{ineq-first-proof-2015}.

\appendix

\section{Lemmata}

\begin{lemma}
\label{yal1}
For a projector $\Pi$, partial isometry $V$ ($V^\dag V = \eye$), Hermitian $X$, and $\delta > 0$,
we have
\begin{align}
\tr \exp\left[ (\Pi V^\dag) \cdot X + (\log \delta) \Pi^\bot \right] & \leqslant \tr \exp\left( X \right) +
\delta \tr \Pi^\bot \\
\left\| \exp \left[ (\Pi V^\dag) \cdot X + (\log \delta) \Pi^\bot \right] \right\| & \leqslant
\max \left[ \left\| \exp \left( X \right) \right\|, \delta \right].
\end{align}
\end{lemma}
\begin{proof}
Let $Z = (\Pi V^\dag) \cdot X$. Let the spectral decompositions be
$X = \sum_i \alpha_i \ketbra{x_i}$ and $Z = \sum_j \beta_j \ketbra{z_j}$. We have
$\sum_j \beta_j \ketbra{z_j} =$ $\sum_i \alpha_i \Pi V^\dag \ketbra{x_i} V \Pi$ or
$\beta_j = \sum_i \alpha_i \Upsilon_{ij}$, where $\Upsilon_{ij} = | \bra{z_j} \Pi V^\dag \ket{x_i} |^2$,
$\sum_i \Upsilon_{ij} = \brackett{z_j}{\Pi}{z_j}$ $\equiv \vartheta_j$ $\in \{0, 1\}$ since $Z$ and $\Pi$
commute, and $\sum_j \Upsilon_{ij} \leqslant 1$.

We now have
$\tr \exp\left[ (\Pi V^\dag) \cdot X + (\log \delta) \Pi^\bot \right] - \delta \tr \Pi^\bot$
$= \sum_j \exp(\beta_j) \vartheta_j =$ \linebreak
$\sum_j \exp\left( \sum_i \Upsilon_{ij} \alpha_i \right) \vartheta_j$
$\leqslant \sum_{i,j} \Upsilon_{ij} \exp(\alpha_i) \vartheta_j
\leqslant \sum_i \exp(\alpha_i)$ $= \tr \exp(X)$,
where the first inequality uses the convexity of $x \mapsto \exp(x)$ when $\vartheta_j=1$,
and is trivially true otherwise. This proves the first claim.
We also have
$\| \exp[ (\Pi V^\dag) \cdot X + (\log \delta) \Pi^\bot]\| =$
$\max \{ \max_j$ $[ \exp(\beta_j) \vartheta_j], \delta \}$
$= \max \{ \max_j [ \exp( \sum_i \Upsilon_{ij} \alpha_i) \vartheta_j], \delta\}$
$\leqslant$ $\max\{ \max_j[ \sum_i \Upsilon_{ij} \exp(\alpha_i) \vartheta_j], \delta\}$
$\leqslant$ \linebreak $\max[ \max_i \exp(\alpha_i), \delta]$ $= \max[ \| \exp(X) \|, \delta]$.
This proves the second claim.
\end{proof}

\begin{lemma}
\label{yal2}
For a completely positive map $\cN$ and nonnegative $\rho$, $\sigma$, $\text{supp}(\rho)
\subseteq \text{supp}(\sigma)$, we have $\text{supp}[\cN(\rho)] \subseteq \text{supp}[\cN(\sigma)]$.
\end{lemma}
\begin{proof}
Note that $\text{supp}(\rho) \subseteq \text{supp}(\sigma)$ $\Leftrightarrow$
$\Pi_\rho \leqslant \Pi_\sigma$. Suppose $\text{supp}[\cN(\rho)] \subsetneq \text{supp}[\cN(\sigma)]$.
Then $\exists$ $\ket{\varphi}$ such that $\tr \cN(\rho) \ketbra{\varphi} > 0$ and
$\tr \cN(\sigma) \ketbra{\varphi} = 0$ or $\tr \Pi_\rho \cN^\dag(\ketbra{\varphi}) > 0$
and $\tr \Pi_\sigma$ $\cN^\dag(\ketbra{\varphi}) = 0$. But this results in a contradiction since
$0 = \tr \Pi_\sigma \cN^\dag(\ketbra{\varphi})$ $\geqslant \tr \Pi_\rho \cN^\dag(\ketbra{\varphi}) > 0$.
\end{proof}

\begin{lemma}
\label{yal3}
For a completely positive map $\cN$, a nonnegative $\rho$, and
any projector $\widetilde{\Pi} \leqslant \Pi_{\cN(\rho)}^\bot$, we have
$\Pi_\rho \cdot \cN^\dag(\widetilde{\Pi}) = \mzero$.
\end{lemma}
\begin{proof}
Since $\tr \cN(\rho) \Pi_{\cN(\rho)}^\bot$ $= 0$, hence, $\tr \Pi_\rho \cN^\dag[\Pi_{\cN(\rho)}^\bot] = 0$.
Now $0 = \tr \Pi_\rho \cN^\dag[\Pi_{\cN(\rho)}^\bot] \geqslant
\tr \Pi_\rho \cN^\dag[\widetilde{\Pi}] \geqslant 0$, and hence equality holds throughout.
The claim follows since $\Pi_\rho \cdot \cN^\dag [\widetilde{\Pi}]$ is a nonnegative matrix with zero trace.
\end{proof}

\section{Alternate proof for strictly positive matrices}
\label{alt-proof}

\begin{theorem}
\label{theorem0}
For a cptp map $\cN$, strictly positive $\sigma$, $\cN(\tau)$, $\cN(\sigma)$, and
$\widetilde{\cR}_{\sigma,\cN}$ as defined in \eqref{yae5}, we have
\beq
\tr \widetilde{\cR}_{\sigma,\cN}(\tau) \leqslant \tr \tau.
\enq
\end{theorem}
\begin{proof}
Let the partial isometry $V^{A \to BE}$ be an isometric extension of $\cN^{A \to B}$. Define the
projector $\hat{\Pi} \equiv \eye^{BE} - V V^\dag$. For any $\varepsilon > 0$, we define
$\sigma_{1,\varepsilon}^{BE} \equiv V \cdot \sigma + \varepsilon \hat{\Pi}$
and $\Upsilon_\varepsilon \equiv \log \cN(\tau) - \log \sigma_{1,\varepsilon}^B$.
Using \eqref{gen-gt}, we now have
\begin{align}
\label{yae2}
\tr \widetilde{\cR}_{\sigma,\cN}(\tau)
& = \tr \exp \left\{ \log \sigma + \cN^\dag \left( \Upsilon_\varepsilon \right)
+ \cN^\dag \left[ \log \sigma_{1,\varepsilon}^B - \log \cN(\sigma) \right] \right\} \\
& \leqslant \tr \exp \left\{ \log \sigma + \cN^\dag \left( \Upsilon_\varepsilon \right) \right\}
\exp \left\{ \cN^\dag \left[ \log \sigma_{1,\varepsilon}^B - \log \cN(\sigma) \right] \right\} \\
\label{yae1}
& \leqslant \tr \exp \left\{ \log \sigma + \cN^\dag \left( \Upsilon_\varepsilon \right) \right\}
\left\| \exp \left\{ \cN^\dag \left[ \log \sigma_{1,\varepsilon}^B - \log \cN(\sigma) \right] \right\} \right\|.
\end{align}
We attack the first term in \eqref{yae1}.
\begin{align}
\tr \exp \left\{ \log \sigma + \cN^\dag \left( \Upsilon_\varepsilon \right) \right\}
& = \tr \exp \left\{ V^\dag \cdot \left[ \log \sigma_{1,\varepsilon}^{BE} + \Upsilon_\varepsilon
\otimes \eye^E \right] \right\} \\
& \leqslant \tr \exp \left\{ \log \sigma_{1,\varepsilon}^{BE} + \left[ \log \cN(\tau) -
\log \sigma_{1,\varepsilon}^B \right] \otimes \eye^E \right\} \\
& \leqslant \tr \int_0^\infty \sigma_{1,\varepsilon}^{BE} \left[ (t \eye^B + \sigma_{1,\varepsilon}^B)^{-1}
\cN(\tau) (t \eye^B + \sigma_{1,\varepsilon}^B)^{-1} \otimes \eye^E \right] dt \\
& = \tr \int_0^\infty (t \eye^B + \sigma_{1,\varepsilon}^B)^{-1} \sigma_{1,\varepsilon}^{B}
(t \eye^B + \sigma_{1,\varepsilon}^B)^{-1} dt \, \cN(\tau) \\
\label{yae3}
& = \tr \tau,
\end{align}
where the first equality follows since for any $\varepsilon \neq 0$, $\log \sigma_{1,\varepsilon}^{BE}
= V \cdot \log \sigma + \log(\varepsilon) \hat{\Pi}$, and hence,
$V^\dag \cdot \log \sigma_{1,\varepsilon}^{BE} = \log \sigma$, and $\cN^\dag(\gamma)
= V^\dag \cdot (\gamma \otimes \eye^E)$, the first inequality uses
Lemma \ref{yal1}, and the second inequality uses \eqref{t-gt}.

We now attack the second term in \eqref{yae1}.
Let $X^B \equiv \trns_E \hat{\Pi}$, and so
$\sigma_{1,\varepsilon}^B = \cN(\sigma) + \varepsilon X^B$.
\begin{align}
\big\| \exp \left\{ \cN^\dag \left[ \log \sigma_{1,\varepsilon}^B - \log \cN(\sigma) \right] \right\} \big\|
& = \left\| \exp \left( V^\dag \cdot \left\{ \left[ \log \sigma_{1,\varepsilon}^B - \log \cN(\sigma) \right]
\otimes \eye^E \right\} \right) \right\| \\
& \leqslant \left\| \exp \left\{ \left[ \log \sigma_{1,\varepsilon}^B - \log \cN(\sigma) \right]
\otimes \eye^E \right\} \right\| \\
& = \left\| \exp \left\{ \log (\sigma_{1,\varepsilon}^B \otimes \eye^E) - \log \left[ \cN(\sigma)
\otimes \eye^E \right] \right\} \right\| \\
& \leqslant \left\| [\cN(\sigma)]^{-1/2} \cdot \sigma_{1,\varepsilon}^B \otimes \eye^E \right\| \\
& = \left\| \eye^B + \varepsilon [\cN(\sigma)]^{-1/2} \cdot X^B \right\| \\
\label{yae4}
& = 1 + \varepsilon \left\| [\cN(\sigma)]^{-1/2} \cdot X^B \right\|,
\end{align}
where the first inequality uses Lemma \ref{yal1},
and the second inequality uses \eqref{gen-gt}.

Using \eqref{yae3} and \eqref{yae4} in \eqref{yae1}, we get
\beq
\tr \widetilde{\cR}_{\sigma,\cN}(\tau)
\leqslant (\tr \tau) \left\{ 1 + \varepsilon \left\| [\cN(\sigma)]^{-1/2} \cdot X^B \right\| \right\}.
\enq
Since this is true for all $\varepsilon > 0$, the claim follows.
\end{proof}

\end{document}